\documentclass[12pt,preprint]{aastex}
\slugcomment{submitted to ApJ on 24 June 2008}
\shortauthors{Geballe et al.}
\shorttitle{CO in T dwarfs}
\def\sss{\scriptscriptstyle}

\def\teff{T_{\rm eff}}

\begin{document}

\title{Spectroscopic Detection of Carbon Monoxide \\ in Two
Late-type T Dwarfs}

\author{T. R. Geballe\altaffilmark{1}, D. Saumon\altaffilmark{2}, D. A.
Golimowski\altaffilmark{3}, S. K. Leggett\altaffilmark{1}, M.S.
Marley\altaffilmark{4}, K. S. Noll\altaffilmark{5}}

\altaffiltext{1}{Gemini Observatory, 670 N. A'ohoku Place, Hilo, HI
96720; tgeballe@gemini.edu}

\altaffiltext{2}{Los Alamos National Laboratory, PO Box 1663, MS F663,
Los Alamos, NM 87545}

\altaffiltext{3}{Department of Physics and Astronomy, Johns Hopkins
University, 3400 North Charles Street, Baltimore, MD 21218}

\altaffiltext{4}{NASA Ames Research Center, MS245-3, Moffett Field, CA
94035}

\altaffiltext{5}{Space Telescope Science Institute, 3700 San Martin
Drive, Baltimore, MD 21218}

\begin{abstract}

$M$ band spectra of two late-type T dwarfs, 2MASS J09373487+2931409, and 
Gliese 570D, confirm evidence from photometry that photospheric CO is 
present at abundance levels far in excess of those predicted from 
chemical equilibrium.  These new and unambiguous detections of CO, 
together with an earlier spectroscopic detection of CO in Gliese 229B 
and existing $M$ band photometry of a large selection of T dwarfs, 
suggest that vertical mixing in the photosphere drives the CO abundance 
out of chemical equilibrium and is a common, and likely universal 
feature of mid-to-late type T dwarfs. The $M$ band spectra allow 
determinations of the time scale of vertical mixing in the atmosphere of 
each object, the first such measurements of this important parameter in 
late T dwarfs. A detailed analysis of the spectral energy distribution 
of 2MASS J09373487+2931409 results in the following values for 
metallicity, temperature, surface gravity, and luminosity: [M/H]$\sim 
-0.3$, $\teff=925-975\,$K, $\log g=5.20-5.47$, $\log L/L_\odot=-5.308 
\pm0.027$. The age is 3--10$\,$Gyr and the mass is in the range 
45--69$\,M_{\rm Jupiter}$.

\end{abstract}


\section{Introduction}

In the 1500--4000~K photospheres of late-type stars and early-mid L-type 
brown dwarfs carbon is predominantly locked up in carbon monoxide (CO), 
due to the much higher binding energy of CO than that of any other 
carbon-bearing molecule. At temperatures of $\sim$1500 K, however, 
methane (CH$_4$) begins to be stable against collisional dissociation. 
As a brown dwarf cools below this temperature, the overwhelming 
abundance of hydrogen slowly drives the carbon that was in CO into 
CH$_4$ \citep{fl96}. This chemical change leads to a striking 
transformation of the infrared spectrum of the brown dwarf and marks the 
division between the L and T classifications 
\citep{kir99,bur02,geb02,bur06}.

Absorption bands of CH$_4$ together with those of H$_2$O dominate the 
1-2.5~$\mu$m spectra of T dwarfs, making the $\lambda$ = 2.3--2.5~$\mu$m 
first overtone band of CO, which is easily observed in K to L dwarfs, 
virtually undetectable by spectral type T4 \citep{geb02}, even if a 
significant amount of CO is still present.  In contrast, the fundamental 
vibration-rotation band of CO is centered near 4.7$\mu$m, where methane 
is not an important absorber. Over a decade ago \citet{noll97} 
discovered evidence of CO at 4.7~$\mu$m at an abundance roughly three 
orders of magnitude larger than expected from chemical equilibrium, in 
the bright T7p brown dwarf Gliese~229B. The discovery was promptly 
confirmed by \citet{opp98}. \citet{noll97} found that the CO abundance 
in Gl~229B matched that expected at a temperature of 1250~K, roughly 
400~K higher than the photospheric temperature.  They interpreted the 
detection as evidence for vertical transport of CO from hotter internal 
atmospheric layers to the surface, which was later supported by more 
detailed analysis \citep{saumon00}.

This finding was not unexpected, as mixing is predicted to occur in the
outer, radiative zones of brown dwarf atmospheres, and the reaction(s)
moving carbon from CO to CH$_4$ are slow \citep{fl94,fl96,lf02}.
Recently, the $M$ band fluxes of numerous late T dwarfs have been
measured to be up to one magnitude fainter than predictions based on
their brightnesses in other wavebands and the assumption of equilibrium
chemistry \citep{saumon03,gol04,patten06}. This strongly indicates that
the CO phenomenon seen in Gl 229B is common to most or all late T
dwarfs. Here we describe spectroscopic observations of two late-type T
dwarfs, 2MASS~09373487+2931409 (hereafter, 2MASS~0937) and Gliese~570D,
whose physical properties differ significantly from Gl~229B, in order to
test this hypothesis spectroscopically.

2MASS~0937 and Gl~229B have nearly the same spectral classifications, 
T6p and T7.5, respectively \citep{bur06}. However, the $K$ band flux of 
2MASS~0937 is significantly depressed relative to Gl~229b and other 
T6-T7 dwarfs; e.g., its $J-K$ color is 0.8~mag bluer \citep{bur02, 
kna04}. This is believed to be due to the effect of greatly enhanced 
collision-induced absorption by H$_2$ (implying that 2MASS~0937 is 
considerably more massive than other T6-T7 dwarfs) and/or that it has 
unusually low metallicity 
\citep{bur02,burgasser03,burrows02,kna04,bbk06}. The $L-M$ color of 
2MASS~0937 \citep{gol04} is close to that of Gl~229B, suggesting that 
its photosphere does contain an enhanced abundance of CO, but a 
spectroscopic observation is needed to verify that the this is indeed 
the cause.

Gl~570D \citep{bur00} is part of a quadruple system whose other members 
are low mass main sequence stars. \citet{saumon06} determined an age for 
it of 3--5~Gyr and a mass of 38-47~M$_{\rm Jupiter}$ by comparing 
evolutionary models of brown dwarfs and 1-2.5~$\mu$m synthetic spectra 
of Gl~570D to the known luminosity and the observed spectrum. Comparison 
of {\it Spitzer} 3.5--7.9~$\mu$m photometry and models strongly suggests 
that the abundance of CO is enhanced by vertical mixing 
\citep{patten06,leg07a}. The {\it Spitzer} 5--15$\,\mu$m spectrum of 
Gl~570D provides strong evidence for mixing also being responsible for 
the depletion of NH$_3$ \citep{saumon06}. In late T dwarfs the nitrogen 
chemistry is insensitive to the time scale of mixing in the radiative 
zone, because it is quenched in the convection zone deep in the 
atmosphere. On the other hand the carbon chemistry, and thus the 
abundance of CO, is very sensitive to the mixing time scale. Since 
Gl~570D is significantly cooler than Gl~229B and 2MASS~0937, 
observations of the 4.7$\,\mu$m band of CO provide an opportunity to 
determine the mixing time scale for a different set of atmospheric 
parameters.

This paper is structured as follows. Section 2 discusses the properties
of the fundamental CO band, \S3 presents the $M$ band spectroscopic
observations and data reduction, and \S4 covers the modeling and
analysis of Gl 570D and 2MASS 0937, respectively.  The last section
summarizes and compares the results for the two objects.

\section{The Spectral Appearance of the Fundamental Band of CO}

In the photospheres of cool stars and ``warm'' brown dwarfs the 
fundamental ($\Delta$$v$~=~1) vibration-rotation band of CO, as viewed 
at low or medium resolution, is a broad and more or less structureless 
absorption stretching across the entire $M$ band. This is the 
consequence of (1) several vibational levels $v$ being populated, (2) 
each $\Delta$$v$~=~1 (i.e., 1--0, 2--1, 3--2, ...) band being 
successively shifted by $\sim$0.06~$\mu$m ($\sim$25~cm$^{-1}$) to longer 
wavelength, and (3) the spacing of vibration-rotation lines in each 
$\Delta$$v$~=~1 band being slightly different. If the rotational levels 
of only the ground vibrational state are populated, however, the 
absorption spectrum of CO observed at low or medium spectral resolution 
will exhibit a bump (due to minimum absorption) around the center of the 
$v=1--0$ band at 4.665~$\mu$m that separates the P and R branches of the 
band. As excited vibrational levels become more populated at higher 
temperatures, this absorption gap gradually disappears.

The absorption gap is predicted to become noticeable at photospheric 
temperatures below 1500~K; that is, at about the same temperature at 
which equilibrium CO abundances are dropping rapidly in the photosphere 
of a cooling brown dwarf. Thus, for a mid-late type (T~$<$~1000~K) T 
dwarf photosphere in chemical equilibrium, CO absorption lines will be 
very weak and no gap will be evident.  If, however, the abundance of CO 
is considerably enhanced, then a broad absorption with a gap at the band 
center should be relatively prominent and detectable at moderate 
spectral resolution. The effect is illustrated in early non-equilibrium 
model spectra of cool dwarfs of \citet{saumon03}.

H$_2$O also is an important absorber in the $M$ band, but its lines,
which are irregularly spaced, tend to be spread across the $M$ band
(although generally increasing in strength toward longer wavelengths).
The H$_2$O lines may partially mask the CO absorption gap at band
center, but if the gap is prominent they should not entirely conceal it.

\section{Observations and Data Reduction}

$M$ band spectra of 2MASS~0937 and Gliese~570D were accumulated over 
several nights during 2004-2005, at the Frederick C. Gillett Gemini 
North telescope, using the facility near infrared imager/spectrograph, 
(NIRI \citet{hodapp03}), which contains grisms for low resolution 
spectroscopy.  The instrument was configured with a $0.75\arcsec$ slit 
and a grism, which together provided a resolving power of 460 
(corresponding to $\Delta$$\lambda$=0.010~$\mu$m) in the $M$ band.

An observing log is given in Table~1. The total exposure times were 
6.1~hr for 2MASS~0937 and 6.5~hr for Gl~570D. Observations were made in 
the standard stare-nod mode so that spectra were obtained alternately at 
two locations along the array, separated by $3\arcsec$. Coaddition of 
subtracted pairs of frames then result in a final sky-subtracted image 
containing both a positive and a negative spectrum of the brown dwarf. 
Positioning of the target in the slit was checked using $J$-band imaging 
and was adjusted if necessary every 45-60 minutes. Typical adjustments 
were $0.1\arcsec$; thus each brown dwarf probably was centered well 
within the slit at all times during most integrations.  A telluric 
standard (a mid F dwarf) was observed just before or just after the 
brown dwarf, and at an airmass that closely matched that of the brown 
dwarf. The seeing, as judged from the intensity profiles along the slit 
of bright calibration stars, ranged from $0.5\arcsec$ to $0.7\arcsec$.

To our knowledge, at $M\sim$12.3 (as determined here) Gl~570D is the
faintest astronomical object to have its $M$-band spectrum successfully
observed from the ground.  Data reduction of $M$ band spectra of faint
sources such as Gl~570D and 2MASS~0937 is difficult, because not only is
the residual background in the sky-subtracted coadded spectral image
considerably brighter than the brown dwarf, but also the gradient in the
residual background along the slit is steep compared to the signal
produced by the spectrum of the faint source, even after $\sim$1~hr of
integration. For these data it was necessary to extract sky rows
adjacent on either side of each of the positive and negative spectra, in
order to remove this gradient, prior to combining the positive and
negative spectra.

Wavelength calibrations of the spectra were obtained from the host of 
strong telluric lines evident in the spectra of telluric standards 
obtained just before and/or after the spectra of the brown dwarfs. The 
accuracies of the calibrations were better than 0.001~$\mu$m 
(3$\sigma$). Flux calibrations utilized the 2MASS $K_S$ magnitudes of 
the telluric standards and the very small color corrections to derive 
their $M$ magnitudes. It was assumed that the spectra of the standards 
and the brown dwarfs suffered identical slit losses due to seeing and 
guiding. That assumption might be suspect, both because the brighter 
stars can be more accurately centered in the slit of the spectrograph 
and because during the longer integration times on the brown dwarfs it 
is more likely that guiding ``drifts'' occur and result in higher signal 
losses.  Nevertheless, the average flux density in the 4.55--4.80~$\mu$m 
spectrum of 2MASS~0937 in Fig.~1, 
5.0~$\pm$0.1~$\times$~10$^{-16}$~W~m$^{-2}$~$\mu$m$^{-1}$, agrees to 
within several percent with the $M'$ (4.55-4.79~$\mu$m) magnitude of 
11.74~$\pm$~0.10) reported by \citet{gol04}. No $M'$ photometry is 
available for Gl~570D. From Fig.~1 we estimate $M'$~=~12.3$\pm$0.3~mag, 
where the uncertainty is largely that of calibration. This value is 
consistent with the \citet{patten06} IRAC [4.5] photometry and values of 
[4.5]-$M'$ for late T dwarfs as given by \citet[][see their 
Fig.~2]{leg07b}; the uncertainty in the color is several tenths of a 
magnitude.

In Figure~1 the coadded, ratioed, and flux-calibrated spectra are binned 
into 0.01~$\mu$m intervals. The spectra are qualitatively similar to the 
published spectra of Gl~229B \citep{noll97,opp98}, but have higher 
signal-to-noise ratios. Both have maximum flux density in the 
4.64-4.70~$\mu$m region, as expected if photospheric CO depresses the 
spectrum everywhere except near the 1-0 band center. The wavelengths of 
the 1-0 CO lines and their relative strengths, assuming optically thin 
lines and LTE conditions for T=750 K, are shown in the upper panel of 
Fig.~1 to illustrate the effect of the band center.  Lines of H$_{2}$O 
also are expected to contribute spectral features. The model spectra 
(see below) show that the strongest H$_{2}$O absorption feature in this 
region is at 4.68~$\mu$m, not far from the CO band center, and indeed 
the spectrum of each brown dwarf shows a pronounced dip at that 
wavelength. Thus the overall shapes of the spectra constitute 
unambiguous detections of CO in both 2MASS~0937 and Gl~570D. The effect 
of the band center is considerably more pronounced in the spectrum of 
Gl~570D than in 2MASS~0937.  In the former the flux density at the 
center is $\sim$50\% above the surrounding, whereas in the latter the 
contrast is roughly half of that.

\section{Modelling and Analysis}

We analyze the $M$ band spectra with the goals of establishing the 
presence of CO on a quantitative basis and determining the time scale of 
vertical mixing in the atmosphere.  The parallaxes of both objects are 
known and red, near-infrared, and mid-infrared {\it Spitzer Space 
Telescope} spectra are available, together sampling about 70\% of the 
total flux emitted. The distance and spectra effectively constrain the 
luminosity, $L$, effective temperature, $\teff$, gravity, $g$, and 
metallicity. The {\it Spitzer} Infrared Spectrograph 
\citep[IRS,][]{houck04} spectra of T dwarfs later than $\sim$T1 show 
NH$_3$ features \citep{cushing06} whose strengths are influenced by 
non-equilibrium nitrogen chemistry and thus by vertical mixing. Those 
features are not sensitive to the time scale of atmospheric mixing, 
however. Once other parameters have been constrained, the strength of 
the CO 4.7$\,\mu$m band can be used to find the mixing time scale. As in 
our previous work \citep{geballe01,saumon06}, we parametrize the 
vertical mixing time scale $\tau_{\rm mix}$ by the eddy diffusion 
coefficient $K_{zz}=H^2/\tau_{\rm mix}$, where $H$ is the mixing scale 
height, assumed to be equal to the local pressure scale height 
\citep{gy99}.

Our analysis is based on the atmospheric models and mixing scheme
described in \citet{saumon06,saumon07} and the evolution sequences of
\citet{sau08}.  Since both Gl~570D and 2MASS~0937 are late T dwarfs, we
assume that their atmospheres are cloudless, which is amply supported by
their $JHK$ photometry and the detailed analysis of the T7p dwarf
Gl~229B \citep{saumon00}, Gl~570D \citep{saumon06}, and the T8 dwarf
2MASS~J0415195$-$093506 \citep{saumon07}.  The evolutionary sequences
were computed with consistent surface boundary conditions provided by a
grid of cloudless atmospheric models with the appropriate metallicities.

\subsection{Gliese 570D}

Because it is a companion to a well-studied main sequence star, Gl~570D
has the most precisely determined physical parameters of any T dwarf to
date. The K4 V primary provides an accurate distance and
metallicity and a tight age constraint on the system. We obtain the
basic parameters from the bolometric luminosity $L$ derived using the
method developed by \citet{geballe01}. Briefly, we use the well-sampled
spectral energy distribution (SED) and a combination of synthetic
spectra and evolutionary sequences to obtain $L$.  Given $L$, the
evolutionary sequences give the range of possible models $\teff(g)$ that
are consistent with the age.  A detailed comparison with the observed
spectrum can further limit the possible range of solutions. This is a
robust method to determine $L$, $\teff$ and $g$ for T dwarfs, as the
results depend only weakly on the metallicity assumed for the models,
the details of the synthetic spectra, and the degree of sampling of the
SED.  The method involves no assumption other than some confidence in
the models, which have been validated by comparison with data
\citep{saumon06,saumon07}.

For Gl~570D, \citet{saumon06} found $\teff=800$--820$\,$K, $\log 
g\,({\rm cm/s}^2)=5.09$--5.23, and $\log L/L_\odot= -5.525$ to $-5.528$. 
The above ranges of values are not due to random uncertainties, because 
the parameters are correlated; e.g., for $\teff=800\,$K, $\log g=5.09$ 
and $\log L/L_\odot=-5.525$.  Fits of the spectrum favor the 
high-$\teff$, high-$g$ limit of this range. The metallicity of the 
system is [Fe/H]=0.09$\pm$0.04 \citep[an average of several 
determinations:][]{tf00,santos05,valf05}. \citet{saumon06} used solar 
metallicity models to obtain the above parameters and we do likewise in 
our analysis of the $M$ band spectrum.

The 5--15$\,\mu$m {\it Spitzer} spectrum shows strong ammonia features
that can be accurately modeled using an NH$_3$ abundance reduced by a
factor of $\sim$10 compared to that in chemical equilibrium, as expected
for non-equilibrium chemistry driven by vertical mixing in the
atmosphere \citep{lf02,saumon06}.  We therefore expect that the $M$ band
spectrum will show a non-equilibrium abundance of CO. We have performed
the analysis using a baseline model computed with the most likely values
determined by \citet{saumon06}, which correspond to their model C:
$\teff=820\,$K, $\log g=5.23$, and an age of 5$\,$Gyr (solar
metallicity). The only free parameter left to fit the $M$ band spectrum
is the eddy diffusion coefficient $K_{zz}$, which determines the surface
abundance of CO.

\subsubsection{Fits of the $M$ band spectrum}

There are two approaches to fitting the spectrum shown in Fig.~1 with a 
set of synthetic spectra.  The first method fits only the shape of the 
spectrum, i.e. the synthetic spectra are scaled by a constant 
multiplicative factor that is adjusted to minimize $\chi^2$ with the 
data.  In this approach, the uncertainty in the fitting parameter 
($K_{zz}$) is determined by the random noise in each data point of the 
observed spectrum. This is estimated at $\sigma=5.0\times 
10^{-17}\,$W$\,$m$^{-2}\,\mu$m$^{-1}$ (Fig.~1). The second 
approach uses evolutionary sequences to obtain the radius of the brown 
dwarf and computes absolute model fluxes that are directly compared to 
the observed fluxes.  This method simultaneously fits the shape of the 
spectrum and its overall flux level and is subject to the systematic 
uncertainty in the flux calibration of the data, which is estimated at 
$\pm30$\%.  We compute $K_{zz}$ and its uncertainty using both methods. 
With perfectly calibrated data and perfect models, both methods would 
give the same value of $K_{zz}$.  We find that the flux calibration 
uncertainty dominates the noise when fitting the absolute fluxes.  The 
final source of uncertainty in the determination of $K_{zz}$ arises from 
the possible range of model parameters for the brown dwarf.  For 
Gl~570D, where $\teff$ is narrowly constrained, this last contribution 
is negligible.

To account for the influence of noise on the interpretation of the 
spectrum, we generated 1000 simulated observed spectra by adding random 
Gaussian noise with the above dispersion to the observed flux in each 
data point. Each of the simulated observed spectra was fitted with a 
synthetic spectrum computed with the baseline model parameters and $\log 
K_{zz}$(cm$^{2}$/s) ranging between 2 and 8.5.  The high resolution 
($\lambda/\Delta \lambda \sim 16000$) synthetic flux densities were 
converted to the resolution of the spectrum in Fig.~1 by integration 
over 0.01~$\mu$m wide bins. The best value of $\log K_{zz}$ is obtained 
by minimizing

\begin{equation} \chi^2 = {1 \over N-3} \sum_{i=1}^N \Bigg[{af_i^{\rm
obs} - \beta f_i^{\rm mod}(K_{zz}) \over \sigma} \Bigg]^2 \end{equation}

\noindent where $N$ is the number of points in the $M$ band spectrum,
$f_i^{\rm obs}$, $f_i^{\rm mod}$, and $\sigma$ are the observed flux,
the absolute model flux, and the random noise in pixel $i$,
respectively. The flux calibration uncertainty of $\pm30$\% enters as
$a$ ($a=0.7$,1, 1.3) and $\beta$ is the model flux renormalization. For
the absolute flux fitting procedure $\beta$ was set to unity. When
fitting the shape of the spectrum only, $\beta$ was adjusted to minimize
$\chi^2$ for each value of $K_{zz}$. The resulting distribution of 1000
values of $K_{zz}$ gives an average value and its dispersion.  The
procedure was repeated for all three values of $a$ (with $\beta=1$) to
quantify the effect of the uncertainty of the flux calibration.

The general behavior of the $\chi^2$ for these different assumptions is 
shown in Fig.~2.  For a given value of $K_{zz}$, the best 
fit, indicated by the lowest $\chi^2$ value, occurs when the model 
fluxes are scaled to the data, i.e. when only the shape of the observed 
spectrum is fitted (dotted curve).  This curve has a broad minimum and 
our statistical sampling finds the minimum at $\log K_{zz}{\rm 
(cm}^2/{\rm s)}=6.21 \pm 0.73$.  Equilibrium chemistry ($K_{zz}=0$), 
which \citet{saumon06} found is strongly excluded on the basis of the 
NH$_3$ depletion revealed by the {\it Spitzer} spectrum, is also 
strongly excluded by the shape of the $M$ band spectrum.  This is 
readily seen in Fig.~3 which compares the 
equilibrium model spectrum, which shows no detectable CO, with the 
best-fitting non-equilibrium model spectrum.  Furthermore, by increasing 
the statistical sampling by one order of magnitude to 10,000 simulated 
observed spectra, we found only three simulated spectra giving $\log 
K_{zz}\,<4$ and none below 3.8.  This illustrates the negligible 
likelihood that the model using chemical equilibrium is consistent with 
the data.  If we fit the data by considering the absolute flux 
calibration, we find narrower distributions of $\chi^2$ (solid curves on 
Fig.~2 that depend sensitively on the flux calibration of 
the data. For the nominal calibration, we get $\log K_{zz}\,=3.98 \pm 
0.12$. Decreasing the absolute flux level by the estimated uncertainty 
of 30\% we find $\log K_{zz}\,=5.23 \pm 0.13$.  The systematic 
uncertainty in $\log K_{zz}$ due to flux calibration is $\sim$8 times 
larger than that due to random noise in the spectrum. Consistent 
solutions between the two fitting methods can be obtained if the 
calibrated flux density is reduced by a factor of 0.65 to 0.5, 
corresponding to a -1.1$\sigma$ to -1.7$\sigma$ correction, which is 
just plausible.

\subsubsection{Photometry}

Photometric measurements of Gl 570D in the MKO $L^\prime$ \citep{gol04} 
and the {\it Spitzer} IRAC [3.6] and [4.5] bands \citep{patten06} 
provide additional relevant information.  We generated synthetic fluxes 
by integrating the model spectra over the bandpasses\footnote{MKO 
system: \citet{tokun02}; IRAC: 
http://ssc.spitzer.caltech.edu/irac/spectral\_ response.html}, and 
multiplying by $(R/D)^2$, where $R(\teff,g)$ is the radius obtained from 
our cloudless [M/H]=0 evolution sequences and $D$ is the distance 
\citep{perryman97}.  The observed magnitudes were converted to fluxes by 
computing zero magnitude fluxes in each bandpass with a spectrum of 
Vega\footnote{www.stsci.edu/instruments/observatory/cdbs/calspec} or 
from the flux calibration of the IRAC 
instrument\footnote{http://ssc.spitzer.caltech.edu/irac/calib/}.  A 
least squares fit of the three photometric points gives $\log 
K_{zz}\,{\rm(cm}^2{\rm /s)}=4.49 \pm 0.24$ for our nominal model 
($\teff=820\,$K, $\log g=5.23$ and [M/H]=0).  This best-fitting 
synthetic spectrum, its corresponding equilibrium spectrum and the 
photometry are shown in Fig.~4 along with the $M$ band spectrum. The $M$ 
band spectrum in the figure is scaled by a factor of 0.83 relative to 
the flux-calibrated spectrum in order to minimize the residuals with the 
model spectrum.  This is a shift of about half of the calibration 
uncertainty of the spectrum. The figure clearly shows the strong and 
broad CO band, centered at 4.7$\,\mu$m and complete with the bump at 
band center, that appears when vertical transport drives the carbon 
chemistry far out of equilibrium. The spectrum outside of the 
4.4--5.0~$\mu$m region is barely affected, however.

\subsubsection{Summary: Optimal model}

Figure~4 shows that a cloudless model with
$\teff=820\,$K, $\log g=5.23$, [M/H]=0 and $\log K_{zz}\,{(\rm cm}^2{\rm
/s)}= 4.5$ agrees with (1) the optical, near infrared and the
5--15$\,\mu$m {\it Spitzer} IRS spectrum \citep{saumon06}, (2) the MKO
$L^\prime$ and {\it Spitzer} IRAC [4.5] photometry, and (3) the flux
level of the $M$ band spectrum within its uncertainties. The IRAC [3.6]
flux is too low, however. In addition $\log K_{zz}\,=6.21 \pm 0.73$
provides a better fit to the shape of the spectrum.  This latter value
also fits all of the data except for the photometry.  Within the various
sources of uncertainty, these two results for $\log K_{zz}$ are barely
compatible. For instance, the (non-Gaussian) statistical distribution
that led to the determination $\log K_{zz}\,=6.21\pm0.73$ has only 1.6\%
of the sample falling within the range $4.5\pm0.24$.

We believe that fitting the shape of the $M$ band spectrum only is the 
more reliable way to determine the value of $K_{zz}$. This method 
depends primarily on the CO/H$_2$O abundance ratio in the photosphere 
and much less on modeling details While the flux level in the {\it 
Spitzer} 4.5~$\mu$m band is indicative of CO absorption, it does not 
specifically point to CO as the absorber and the filter is not well 
matched to the CO band. In addition, there is no photometric data point 
outside of and just to longer wavelength of the CO band, where 
absorption by H$_{2}$O, which also occurs within the CO band, should 
dominate.

The photometric measurements are much more precise than the $M$ band 
spectrum calibration, however.  They measure the spectral energy 
distribution in the vicinity of the 4--5$\,\mu$m peak, which is strongly 
affected by the $\nu_3$ band of CH$_4$ centered at 3.3$\,\mu$m extending 
to $\sim 4\,\mu$m, and by the $\nu_2$ band of H$_2$O on the long 
wavelength side of the peak. The CH$_4$ line list is known to be 
incomplete at the temperatures of brown dwarf atmospheres. The modest 
success in fitting the 3--4$\,\mu$m spectra of T dwarfs 
\citep{stephens08} suggests that the inconsistency between the two 
solutions could be resolved in favor of the higher value of $K_{zz}$ by 
invoking a higher CH$_4$ opacity. Indeed the incompleteness of the 
CH$_4$ line list implies that the CH$_4$ opacity in our model is a lower 
limit to the actual opacity.  Increasing the opacity in the red wing of 
the $\nu_3$ band of CH$_4$ would nudge all three broad band fluxes in a 
direction that would increase the value of the best fitting $K_{zz}$. 
The flux removed from the increased CH$_4$ absorption in the near 
infrared will reemerge primarily in low opacity windows.  It is very 
likely that the shape of the 4--5$\,\mu$m peak will be altered by flux 
redistribution caused by a change in the CH$_4$ opacity, which could 
significantly change the value of $K_{zz}$ that best fits the 
photometry.

Thus, we adopt $\log K_{zz}\,{\rm(cm}^2{\rm /s)}= 6.2 \pm 0.7$ as the 
most likely value for Gl~570D, based on the favored nominal model 
parameters (model C of \citet{saumon06}).  Repeating the above analysis 
with the less favored but still possible values of model B of 
\citet{saumon06} ($\teff=800\,$K, $\log g=5.09$), gives $\log 
K_{zz}\,{\rm cm}^2{\rm /s}=6.0 \pm 0.7$ which is essentially identical 
within the uncertainty. The corresponding time scale of vertical mixing 
in the radiative region of the atmosphere of Gl~570D is $\sim 0.1$ to 3 
hours. The complete infrared synthetic spectrum of our optimal model for 
Gl~570D is not shown because it is essentially indistinguishable from 
that shown in Fig. 2 of \citet{saumon06} which was computed with $\log 
K_{zz}\,=2$. The only region of the spectrum that is affected by the 
larger value of $K_{zz}$ is between 3.8 and 5.0$\,\mu$m, which is shown 
in Figs.~3 and~4.

For the optimal model the composition of the atmosphere as a function of 
depth for the most important opacity sources is shown in Fig.~5. 
The chemistry of nitrogen is quenched in the 
convection zone where $\tau_{\rm mix}$ becomes shorter than $\tau_{{\rm 
N}_2}$, the time scale for conversion of N$_2$ to NH$_3$.  This occurs 
at $\log T \sim 3.38$. In the upper atmosphere, the NH$_3$ abundance is 
nearly a factor of ten lower than the equilibrium value. Because the 
eddy diffusion coefficient $K_{zz}$ determines the mixing time scale in 
the {\it radiative} zone, the quenching level of the nitrogen chemistry 
is unaffected by the choice of $K_{zz}$; rather it is determined by the 
convective mixing time scale.  On the other hand, the carbon chemistry 
is quenched in the radiative zone around $\log T=3.17$ and its abundance 
depends sensitively on $K_{zz}$ because the equilibrium mole fraction of 
CO has a steep dependence on depth in the atmosphere in low-$\teff$ 
models. In the upper part of the atmosphere the CO mole fraction is 
$\log X_{\rm CO}=-4.10\pm0.18$.  Finally, the 4.7$\,\mu$m band of CO and 
the 10--11$\,\mu$m feature of NH$_3$ probe the same level of the 
atmosphere, because the lines in both spectral regions are formed at 
$\log T~\sim~2.75$--2.9.

\subsection{2MASS 0937}

\subsubsection{Determination of $\teff$, gravity and metallicity}

Due to its very blue $J-K$ color, 2MASS 0937 has long been recognized as
having low metallicity and/or high gravity
\citep{bur02,burgasser03,burrows02,kna04,bbk06}.  Previous estimates give
 $\teff$~=~725-1000$\,$K \citep{gol04} and 700-850$\,$K \citep{vrba04}, 
and $\log g\sim~5.5$ \citep{kna04}.  More recently \citet{bbk06}, using 
a calibrated set of spectral indices, found that 2MASS~0937 is metal 
poor with [M/H] between $-0.4$ and $-0.1$.  For [M/H]$=-0.2$, they 
obtain $\teff=780$-840$\,$K and $\log g=5.3$-5.5.  With the wealth of 
information provided by the wide spectral coverage now available, we can 
determine these parameters more reliably. The observed spectrum covers 
the intervals 0.63--1.01$\,\mu$m, 1.03--1.345$\,\mu$m, 
1.40--2.53$\,\mu$m and 5.13--15.35$\,\mu$m.  The integrated observed 
flux is $2.97 \times 10^{-12}\,$erg\,s$^{-1}$\,cm$^{-2}$, which 
represents about 70\% of the total flux. The flux calibration 
uncertainties are $\pm3$\% in the red, $\pm5$\% in the near infrared, 
and $\pm3.7$\% for the {\it Spitzer} IRS spectrum.  The distance is 
$6.141 \pm 0.146\,$pc \citep{vrba04}. The resulting uncertainty in the 
bolometric luminosity is $\pm 6.4$\%, or 0.027 dex, where we have 
combined the various flux calibration uncertainties rather than treating 
them as statistically independent variations.

We first applied our suite of solar metallicity models and found, as 
expected, that the model overestimate the $K$ band flux by about 60\%, 
even at the highest possible gravity of $\log g=5.43$. A better match of 
the near-infrared spectrum, and the $K$ band flux in particular, thus 
requires a sub-solar metallicity. Applying our grid of model 
atmospheres, spectra, and evolution for [M/H]=$-0.3$ to the method 
described in \citet{geballe01} and \citet{saumon06}, we obtain the 
parameters shown in Fig.~6 and given in Table 2 
for four different ages ranging from 1 to 10$\,$Gyr. We will see below 
that models corresponding to ages of 1 Gy and less do not fit the data 
satisfactorily. We find that $\log L/L_\odot \sim -5.30$, the same 
luminosity as the T7.5 dwarf 2MASS J12171110$-$0311131 \citep{saumon07}. 
The red and near infrared spectra of these four models are nearly 
identical, as can be seen in Fig.~7 where the two 
extreme cases, models A and D, are plotted.  The lower gravity model 
matches the data best, but the four models reproduce the $J$ and 
$K$ peak fluxes equally well.  Note that the fluxes in the model spectra 
in Fig.~7 are absolute and have not been scaled 
to the observed spectrum. The model spectra were computed from 
parameters determined with the integrated spectral flux and the distance 
to 2MASS~0937 as the only input.  The excellent agreement implies that 
the metallicity of 2MASS~0937 is close to [M/H]$=-0.3$, the value used 
for the rest of the analysis.  For comparison, the values obtained with 
the solar metallicity models are shifted by 10-15$\,$K lower in $\teff$, 
$\sim 0.05\,$dex lower in $\log g$ and $<0.02\,$dex lower in $\log L$ 
for a given age. These small variations, which are close to the 
uncertainties in the present determinations, demonstrate that a more 
accurate determination of the metallicity of 2MASS 0937 will not 
significantly affect the parameters given in Table~2.

The range of parameters between models A and D cannot be further 
constrained on the basis of the near infrared spectrum. Generally, the 
$K$ band flux (e.g., $J-K$) is a gravity indicator {\it for a fixed} 
$\teff$.  The models in Table~2 are nearly constant in $L$. Higher 
gravities also imply higher $\teff$, with compensating effects on the 
near infrared spectra (Fig.~7). As we had found in 
our analysis of Gl 570D \citep{saumon06}, the mid-infrared {\it Spitzer} 
IRS spectrum is more sensitive to variations in $\log g$ in the suite of 
$\teff(g)$ solutions than is the near-infrared spectrum. We have 
compared each of the four models in Table 2 with the mid-infrared data, 
taking into account the spectrum of random pixel noise and the 
systematic $\pm 3.7$\% calibration uncertainty \citep{patten06}. Because 
there is a strong suspicion that the chemistry in the atmosphere of 
2MASS~0937 is driven out of equilibrium by vertical mixing, we also 
consider out-of-equilibrium synthetic spectra computed with the 
parameters of models A through D.  The mid-infrared spectrum is 
insensitive to the choice of the vertical mixing parameter $K_{zz}$ 
\citep{saumon06} and we arbitrarily choose a value of $10^4\,$cm$^2$/s.  
To constrain the set of $\teff(g)$ solutions, we thus consider eight 
model spectra (the four models of Table~2, with both $K_{zz}=0$ and 
$10^4\,$cm$^2$/s) and the observed 5-15$\,\mu$m spectrum with its 
nominal calibration, the calibration modified by $\pm3.7$\%, and finally 
by fitting only the shape of the spectrum, i.e. without regard of the 
absolute flux level.

The goodness of fit of a particular model is measured by the $\chi^2$, 
which includes the random uncertainty in the flux of each pixel. The 
ensemble of $\chi^2$ for these models reveals a clear picture.  In every 
case, the best of the equilibrium models (A-D) is a much worse fit than 
the best non-equilibrium model, at more than an 8$\sigma$ level of 
significance.  The IRS spectrum thus rules out equilibrium chemistry for 
nitrogen in 2MASS~0937.  For the non-equilibrium case, models A through 
D fit the shape of the spectrum equally well.  If absolute fluxes are 
considered, for all choices of calibration (within $\pm 3.7$\%) the 
lowest gravity model (A) is ruled out at $>22\sigma$. Models B through D 
can provide equally good fits depending on the choice of absolute 
calibration.  For the nominal calibration, the best fit is obtained with 
model C.  We conclude that the parameters of 2MASS~0937 fall within the 
range between models B and D in Table 2 and that vertical transport 
occurs in its atmosphere, resulting in a depletion of NH$_3$. The 
synthetic spectrum for non-equilibrium model C is shown in Fig.~8
as well as the best equilibrium model (model B). 
The non-equilibrium model agrees very well with the data, although the 
fit is not perfect.

To summarize, combining the information in the red, near
infrared, and mid-infrared spectra with the distance to 2MASS~0937 we
find that $\teff=925$ to $975\,$K, $\log g= 5.19$ to 5.47, $\log
L/L_\odot=-5.31 \pm 0.027$, the metallicity is close to [M/H]$=-0.3$,
and the age is between $\sim$3 and 10~Gyr.  These values confirm that
2MASS~0937 is a metal-poor brown dwarf with a fairly high gravity. The
metallicity and gravity agree with the determinations of \citet{bbk06}
and \citet{kna04} but $\teff$ is substantially higher than previous
determinations, except for that of \citet{gol04}.

\subsubsection{Fits of the $M$ band spectrum}

We fit the $M$ band spectrum of 2MASS~0937 to determine the value of the 
eddy diffusion coefficient $K_{zz}$ using the procedure described in 
\S4.1.1. In this case, the random noise in each pixel of the observed 
spectrum is estimated at $\sigma=4.2\times 
10^{-17}\,$W$\,$m$^{-2}\,\mu$m$^{-1}$ and the systematic uncertainty in 
the flux calibration is estimated at $\pm30$\%\ although the agreement 
with previous $M'$ photometry is much better than this.  We consider the 
three models B though D in this analysis. By fitting the shape of the 
$M$ band spectrum only, we find that $\log K_{zz}\,{\rm(cm}^2{\rm 
/s)}=4.3 \pm 0.3$ for model C.  The value increases slowly with $\log g$ 
among the suite of models but the differences with the above result 
remain well within the uncertainty. All three models give equally good 
fits to the shape of the $M$ band spectrum.  An example of such a fit is 
shown in Fig.~9 where the non-equilibrium 
spectrum corresponding to model C with $\log K_{zz}\,=4.3$ is in much 
better agreement with the shape of the $M$ band spectrum than the same 
model in equilibrium.  The latter fails to produce the 4.665$\,\mu$m 
bump seen in the data that indicates the presence of CO.  This 
corroborates our conclusion based on the fit of the 5-15$\,\mu$m 
spectrum that vertical mixing in the atmosphere of 2MASS~0937 causes the 
chemistry to depart from equilibrium.

By fitting absolute model fluxes to the data we obtain lower values of
$\log K_{zz}$ (Table 2).  If we rescale the observed fluxes by factors
of 0.8--0.75 (less than the nominal calibration uncertainty), then
models B through D can all be made to give $\log K_{zz}\,{\rm (cm}^2{\rm
/s)}=4.3 \pm 0.3$, and a consistent solution is found between fitting the
shape of the $M$ band spectrum and its flux level.  This value of the
eddy diffusion coefficient corresponds to a mixing time scale of $\sim
5$ to 25 hours and a CO mole fraction of $\log X_{\rm CO}=-5.10 \pm
0.13$.

\subsubsection{Photometry}

We attempted to fit the MKO $L^\prime$ and $M^\prime$ magnitudes 
\citep{gol04} and the {\it Spitzer} IRAC [3.6] and [4.5] fluxes 
\citep{patten06} as in \S4.1.2. In this case however, it was not 
possible to closely match the shape of the 3.5--5$\,\mu$m peak with any 
of our models, not even with discarded model A (Table~2), for any value 
of $K_{zz}$.  In all cases, the modelled 3.3$\,\mu$m band of CH$_4$ is 
too deep, and the resultant $L^\prime$ and [3.6] fluxes are too low 
(Fig.~10). Values of $K_{zz}$ that fit the [4.5] or 
the $M^\prime$ fluxes can be found but are different from each other.  
Using the model C parameters, we find $\log K_{zz}\,{\rm(cm}^2{\rm 
/s)}=3.3 \pm 2.2$, but both the fit and $\chi^2$ are very poor, with the 
[3.6] flux falling several standard deviations below the IRAC 
measurement.  The 3.3$\,\mu$m band of CH$_4$ can be weakened by further 
decreasing the metallicity, or increasing $K_{zz}$, but the latter to a 
value that is incompatible with the shape of the $M$ band spectrum.  
Both of these possibilities appear unlikely, given the good agreement 
that we find with the spectroscopic data.  A more complete line list for 
the CH$_4$ band would increase the opacity and thus go in the wrong 
direction. The discrepancy between our best fitting model and the [3.6] 
and $L^\prime$ photometry might be attributable to the secondary effect of 
increasing the CH$_4$ opacity and redistributing flux in the low opacity 
windows at wavelengths less than $1.7\,\mu$m into the 4--5$\,\mu$m.

\subsubsection{Summary: Optimal model}

In view of the difficulty in fitting the 3--5$\,\mu$m photometry and the 
likelihood that the model spectra are impacted by an incomplete CH$_4$ 
linelist, we again rely on the shape of the $M$ band spectrum to 
determine the most likely value of $K_{zz}$, and thus conclude that 
$\log K_{zz}\,{\rm(cm}^2{\rm /s)}=4.3 \pm 0.3$.  With the data in hand, 
we are unable to distinguish between models B, C, and D, as they give 
essentially equally good -- or in the case of the photometry, equally 
poor -- fits to the ensemble of data.  Model C ($\teff=950\,$K, $\log 
g=5.35$, [M/H]=$-0.3$) with $\log K_{zz}\,=4.3$ as an example, provides 
an excellent fit to the red and near infrared spectra and to the {\it 
Spitzer} IRS spectrum as well (Fig.~11). It also gives a good fit to the 
shape of the $M$ band spectrum (Fig.~9, although it underestimates the 
$M^\prime$ flux by $1.4\sigma$ (Fig.~10).  This model also 
underestimates the [3.6], [4.5], and $L^\prime$ fluxes substantially, 
however.

The chemical composition of the atmosphere as a function of depth is
shown in Fig.~12. In the region of formation of the
4.7$\,\mu$m band of CO ($\log T=2.9-2.98$), vertical transport increases
the CO abundance by about two orders of magnitude. The NH$_3$ abundance
decreases by $\sim 0.8\,$dex which accounts for the large effect seen in
Fig.~8. The H$_2$O and CH$_4$ mole fractions are
barely affected by non-equilibrium chemistry.  As was found for Gl~570D,
the nitrogen chemistry is quenched deep in the convection zone of the
atmosphere (at $\log T =3.38$) and thus is unaffected by the value of
$K_{zz}$.  On the other hand, the carbon chemistry is quenched much
higher in the atmosphere, at $\log T=3.12$.

\section{Conclusions}

Low resolution $M$ band spectra of Gl 570D and 2MASS 0937 clearly reveal 
the presence of photospheric CO via the CO absorption minimum between 
the P and R branches of the $v=1--0$ band. These are the first 
spectroscopic detections of CO in late T dwarfs since the original 
discovery in the T7p dwarf Gliese 229B. The observed spectra are 
well-reproduced by model spectra with CO mole fractions of $\log X_{\rm 
CO}=-4.10 \pm 0.18$ and $-5.10 \pm 0.13$, respectively.

Atmospheric models and synthetic spectra based on equilibrium chemistry
predict that the CO abundance should be too low to form a detectable CO
band and that instead the $M$ band spectrum should be dominated by
H$_2$O features. Thus, the detection of prominent CO absorption in the
spectra of these two T dwarfs is strong evidence for the presence
of vertical mixing in the radiative part of their atmospheres. This
interpretation is reinforced by independent evidence for depletions of
NH$_3$ by a factor of 8 to 10 compared to the expectations from chemical
equilibrium calculations in both Gl~570D \citep{saumon06} and 2MASS~0937
(\S 4.2.1).  The concordance of deviations of the NH$_3$/N$_2$ and the
CO/CH$_4$ ratios from chemical equilibrium in both objects can be
explained consistently by a model including vertical mixing in the
atmosphere and the kinetics of carbon and nitrogen chemistry. The
present work firmly establishes this process in cool T dwarfs.

The model spectra that include vertical mixing reproduce the entire 
observed SEDs of both dwarfs very well, with the exception of the 
photometry of the 4--5$\,\mu$m peak in 2MASS~0937. Similar difficulties 
have been encountered with 2MASS J04151954$-$0935066 (T8) and 2MASS 
J12171110$-$0311131 (T7.5) \citep{saumon07}.  Given the overall success 
of our cloudless models at reproducing the SEDs of late T dwarfs, we 
think that this problem is related to well-known remaining flaws in the 
models.  Since the 4--5$\,\mu$m peak occurs in a low opacity window, it 
should be sensitive to the flux redistribution that we expect will occur 
in models that will benefit in the future from an improved line list for 
CH$_4$ at wavelengths below 1.7$\,\mu$m (see Fig.~7 for example). Our 
non-equilibrium spectra are currently computed as post-processing on a 
fixed atmospheric structure computed with abundances from chemical 
equilibrium. A self-consistent calculation of non-equilibrium models 
would also redistribute the flux, but the effect should be relatively 
modest \citep{hb07}.

The values of $K_{zz}$ and of $\tau_{\rm mix}$ derived for Gl~570D and
2MASS~0937 depend directly on the choice of reaction pathway for the
kinetics of the carbon chemistry and on the choice of quenching scheme.
As in our previous work, we have adopted the ``fast'' reaction rate of
\citet{yung88} and the quenching scheme of \citet{smith98}.  The latter
is physically and computationally more realistic than the simple scheme
that is generally applied \citep{hb07,fl94}.  Its application to
modeling the excess CO in Jupiter's troposphere then requires the fast
reaction rate of \citet{yung88} to reproduce the spectroscopic data
\citep{bezard02}. We have used those same choices for the kinetics of
carbon chemistry and the quenching scheme in the past, so the present
determinations of $K_{zz}$ and our previous estimates \citep{saumon07}
are consistent.  Using the ``slow'' kinetic scheme of \citet{pb77} would
result in significantly lower values of $K_{zz}$ and longer mixing time
scales.

The most direct way to determine the vertical mixing time scale
$\tau_{\rm mix}$ in the atmosphere with the data currently available is
from the CO abundance derived by fitting models to the shape of the $M$
band spectrum. The time scale is parametrized by the eddy diffusion
coefficient, $K_{zz}=H^2/\tau_{\rm mix}$, for which we find $\log
K_{zz}\,{\rm (cm}^2{\rm /s)}=6.2\pm 0.7$ for Gl~570D, which corresponds
to a mixing time scale of 0.1 to 3 hours. While the uncertainty in the
mixing time scale remains large, it is crudely comparable to the
measured rotation period of $P=3.6\sin i$~hours, where $i$ is the
inclination of the axis of rotation to the line of sight \citep{zo06}.
For 2MASS 0937, $\log K_{zz}\,{\rm (cm}^2{\rm /s)}=4.3 \pm 0.3$,
implying a much longer mixing time scale of $\tau_{\rm mix}=5$ to 25
hours, which is also comparable to the $\sim2$--10 hour rotation periods
of T dwarfs \citep{zo06}. This suggests that the vertical mixing process
is coupled to the rotation. The eddy diffusion coefficients of these two
dwarfs differ by nearly two orders of magnitude, which seems a rather
large variation for two T dwarfs whose physical properties are fairly
similar: $\Delta \log g <0.25\,$, and $\Delta \log L=0.22\,$. Other
properties in which there are significant differences between the two
dwarfs in this study are the effective temperature, which differs by
100--150$\,$K and metallicity, where $\Delta$[M/H]=0.3 to 0.4.  In the
absence of a physical model for vertical mixing in brown dwarf
atmospheres, it is difficult to appreciate the significance of these
differences between Gl~570D and 2MASS~0937.  In this context, a measure
of the rotational velocity of 2MASS~0937 would be of interest.

Eddy diffusion coefficients (or bounds on their values) have been 
derived for the atmospheres of all of the solar system giants, often by 
comparing models with measurements of the vertical profiles of various 
photochemical products. Other methods have been used as well. As 
reviewed by \citet{moses04}, estimates for Jupiter's upper stratosphere 
cluster around $\log K_{zz}\,$(cm$^2$/s)=6, comparable to or larger than 
the values we derive here for T dwarfs, despite Jupiter's four orders of 
magnitude smaller heat flow and nearly two orders of magnitude lower 
gravity.  Constraints have even been placed on eddy mixing in the 
atmosphere of Neptune, where similar values of $\log K_{zz}$ are 
inferred \citep{bishop95}.  Such mixing in the radiatively-stable solar 
system stratospheres is often attributed to the propagation and/or 
breaking of atmospheric waves.  Bishop et al. review the literature in 
some detail and point out that $K_{zz}$ likely varies vertically through 
an atmosphere, increasing with falling density. Such a shortening of the 
mixing time scale with height would leave the quenching level and the 
resulting chemistry unchanged (Figs.~5 and 12). As such, the $K_{zz}$ 
values we derived from the observations are local values at the 
quenching level of the atmosphere, which is at both higher pressures and 
higher densities than the regions typically probed in giant planet 
atmospheres. The relatively similar mixing coefficients in such a great 
diversity of atmospheres is striking, even though the derived values 
apply to different atmospheric regions.  Further investigation of the 
underlying mechanisms responsible for mixing in giant planet and brown 
dwarf atmospheres would be of interest and may also find applicability 
to the study of extrasolar giant planet stratospheres.

Finally, we have derived the first set of well-constrained physical
parameters for 2MASS 0937 and have firmly established that this T6p
dwarf has a subsolar metallicity and a moderately high gravity, as has
long been suggested.  Our analysis shows that, like other late T dwarfs,
its mid-infrared and $M$ band spectra can only be fit with models in
which both the nitrogen and the carbon chemistry depart substantially
from chemical equilibrium. This reinforces the conjecture that vertical
mixing and the associated departures from equilibrium chemistry are
common, if not ubiquitous, among late T dwarfs.

\begin{acknowledgements}

The data presented here were obtained at the Gemini Observatory, which
is operated by the Association of Universities for Research in
Astronomy, Inc., under a cooperative agreement with the NSF on behalf of
the Gemini partnership: the National Science Foundation (United States),
the Science and Technology Facilities Council (United Kingdom), the
National Research Council (Canada), CONICYT (Chile), the Australian
Research Council (Australia), Minist\'erio da Ci\'encia e Tecnologia
(Brazil) and SECYT (Argentina). This work is also based in part on
observations made with the {\it Spitzer Space Telescope}, which is
operated by the Jet Propulsion Laboratory, California Institute of
Technology, under contract with NASA.  Support for this work, part of
the Spitzer Space Telescope Theoretical Research Program, was provided
by NASA. We thank M.C Cushing for providing the {\it Spitzer} IRS
spectra and A.J. Burgasser for the red spectra. This research has
benefited from the M, L, and T dwarf compendium housed at
DwarfArchives.org and maintained by Chris Gelino, Davy Kirkpatrick, and
Adam Burgasser.

\end{acknowledgements}

\clearpage

\begin{figure}
\epsscale{0.8}
\plotone{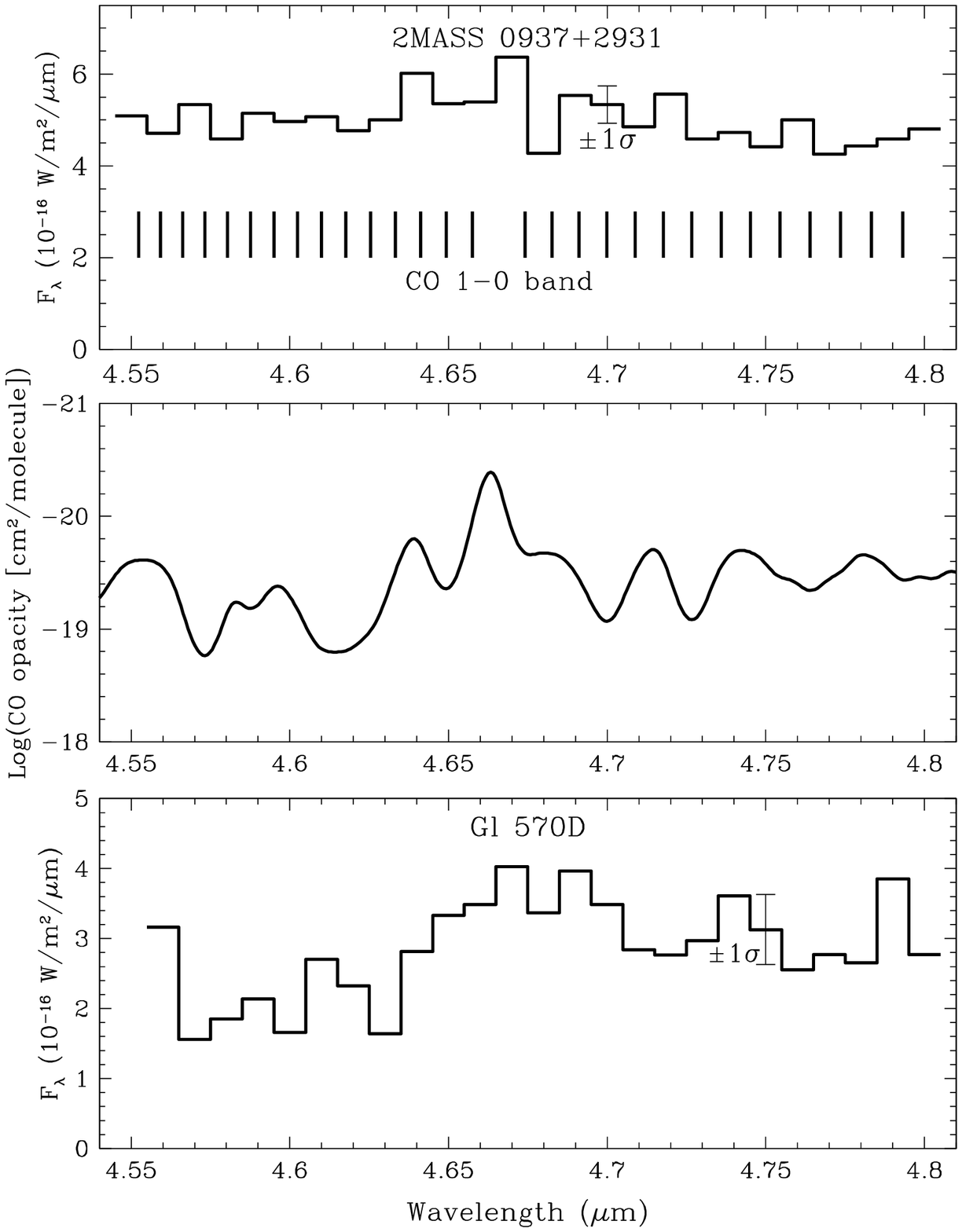}
\caption{Spectra of T dwarfs 2MASS~0937 (upper panel) and
Gl~570D (lower panel) near the CO 1-0 band center. Typical uncertainties
are shown. Wavelengths of the 1-0 band lines and their relative strengths
in optically thin gas at 750~K are shown in the upper panel.}
\label{fig:data}
\end{figure}
\clearpage

\begin{figure}
\plotone{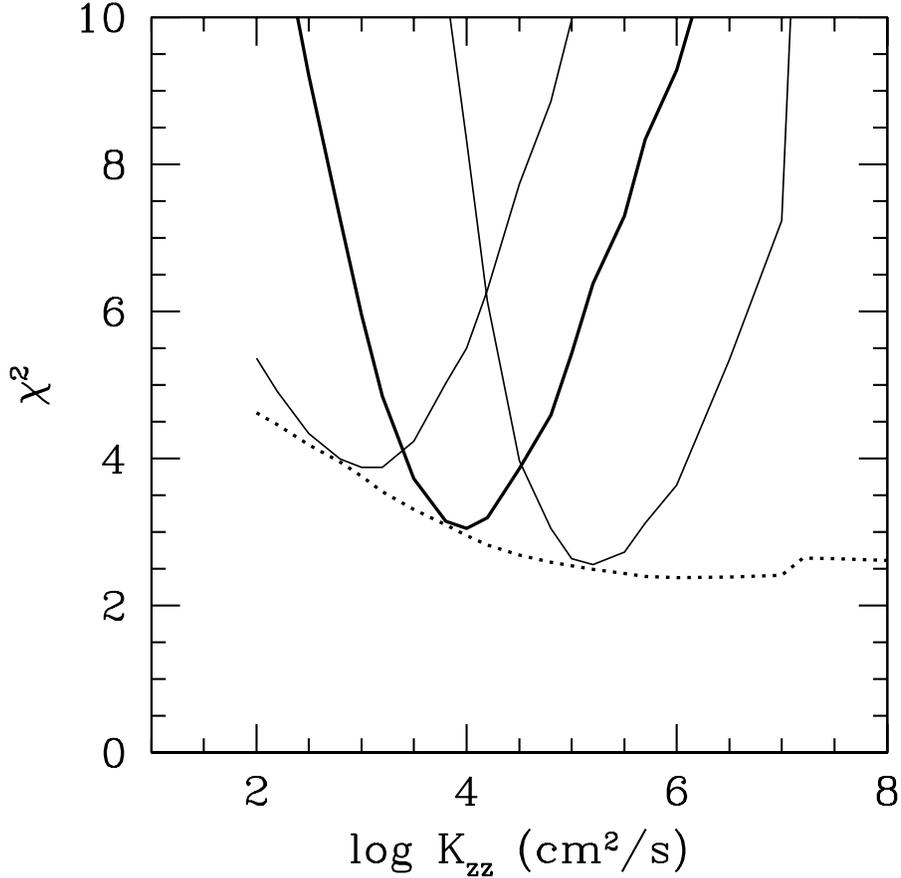}
\caption{Goodness of fit to the $M$ band spectrum as a function of the 
eddy diffusion coefficient $K_{zz}$ for the nominal parameters of 
Gl~570D ($\teff=820\,$K, $\log g=5.23$, [M/H]=0).  $\chi^2$ is defined 
in Eq. 1 (see text).  The dotted curve is obtained when the model 
spectra are scaled to minimize $\chi^2$; i.e. only the shape of the 
observed spectrum is fitted.  The discontinuity in $\chi^2$ near $\log 
K_{zz}$ (cm$^2$/s) =7.2 occurs when the quenching of the carbon 
chemistry goes through the radiative/convective boundary in the 
atmosphere model. The solid curves are obtained when three absolute 
fluxes, corresponding to the nominal calibration (heavy line) and two 
that differ by $\pm30$\%, are included.}
\label{fig:chi2}
\end{figure}
\clearpage

\begin{figure}
\plotone{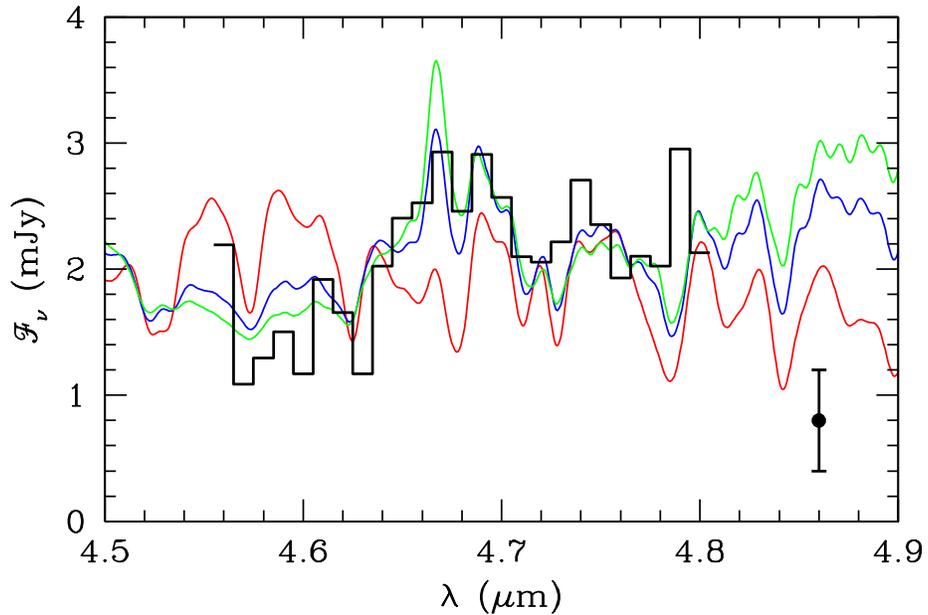}
\caption{Fits to the $M$ band spectrum of Gl 570D. The data are the 
black histogram and the typical pixel noise level ($\pm 1\sigma$) is at 
lower right.  The curves are model spectra plotted at $R$=500 with 
$\teff=820\,$K, $\log g=5.23$, and [M/H]=0 that have been fit to the 
data with scaling factors adjusted to minimize the residuals.  The red 
curve is the equilibrium model ($K_{zz}=0$), the blue curve is the 
non-equilibrium model shown in Fig.~4 ($\log 
K_{zz}\,{\rm cm}^2{\rm /s}=4.5$) and the green curve is the best-fitting 
model ($\log K_{zz}\,{\rm cm}^2{\rm /s}=6.2$). [{\it See the electronic 
edition of the Journal for a color version of this figure.}]}
\label{fig:Mband_fit_Gl570D}
\end{figure}
\clearpage

\begin{figure}
\plotone{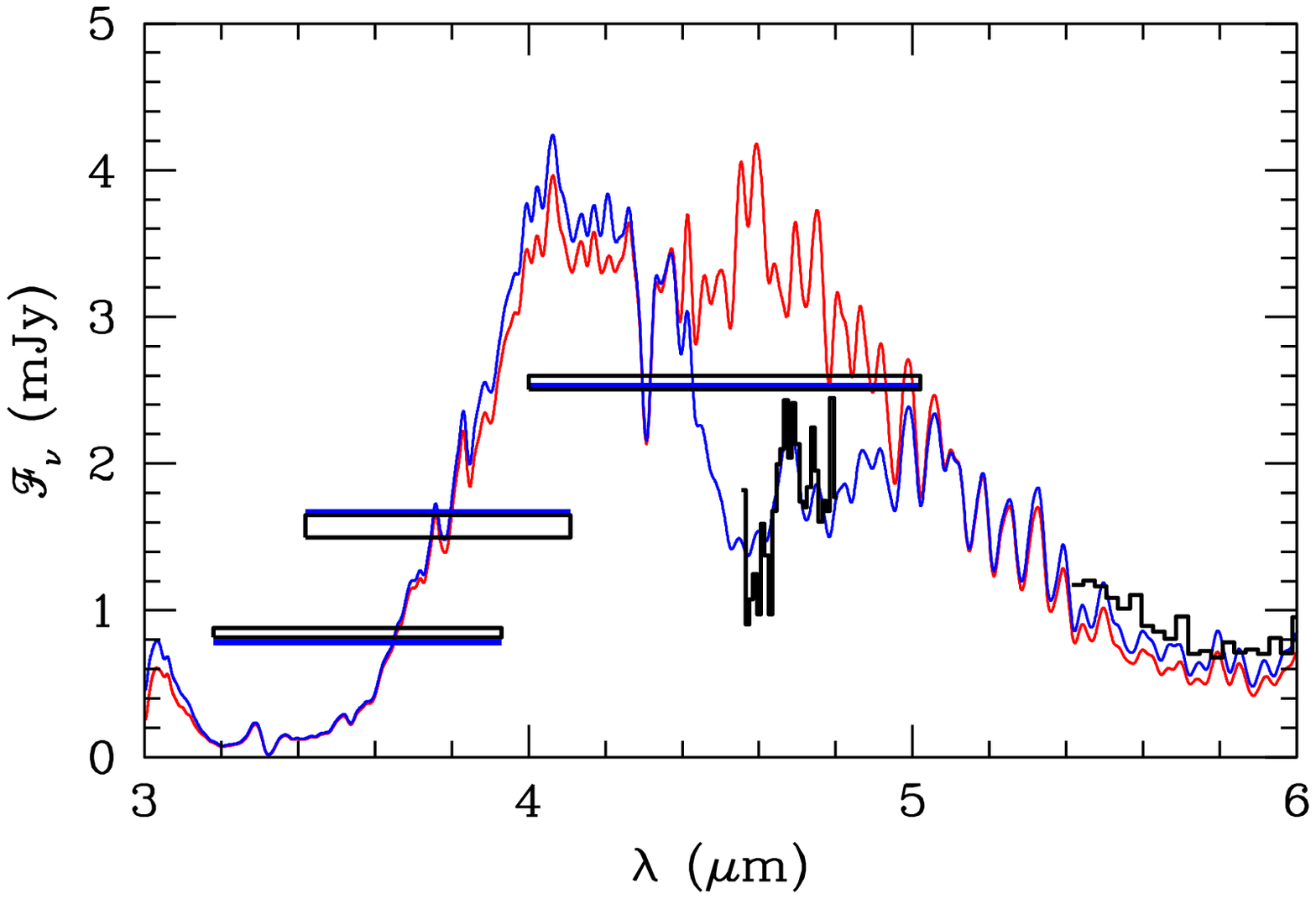}
\caption{3--6~$\mu$m spectral region of Gl 570D.  Data are in
black.  Photometric measurements in the IRAC [3.6], MKO $L^\prime$, and
IRAC [4.5] bands \citep{gol04,patten06} are shown by boxes whose widths
approximate the bandpasses and heights indicate the $\pm 1 \sigma$
uncertainty. Two models are shown at $R=200$, both with $\teff=820$K,
$\log g=5.23$ and [M/H]=0.  The red curve is the equilibrium model
($K_{zz}=0$); the blue curve shows the model that best fits the photometry
($\log K_{zz}\,$(cm$^2$/s)=4.5).  Synthetic models fluxes in the filter
bandpasses are shown by thick blue lines. The observed spectrum has been 
rescaled by a factor of 0.83 to match the model spectrum. [{\it See the 
electronic edition of the Journal for a colorversion of this figure.}]}
\label{fig:sp_Gl570D_2}
\end{figure}
\clearpage

\begin{figure}
\plotone{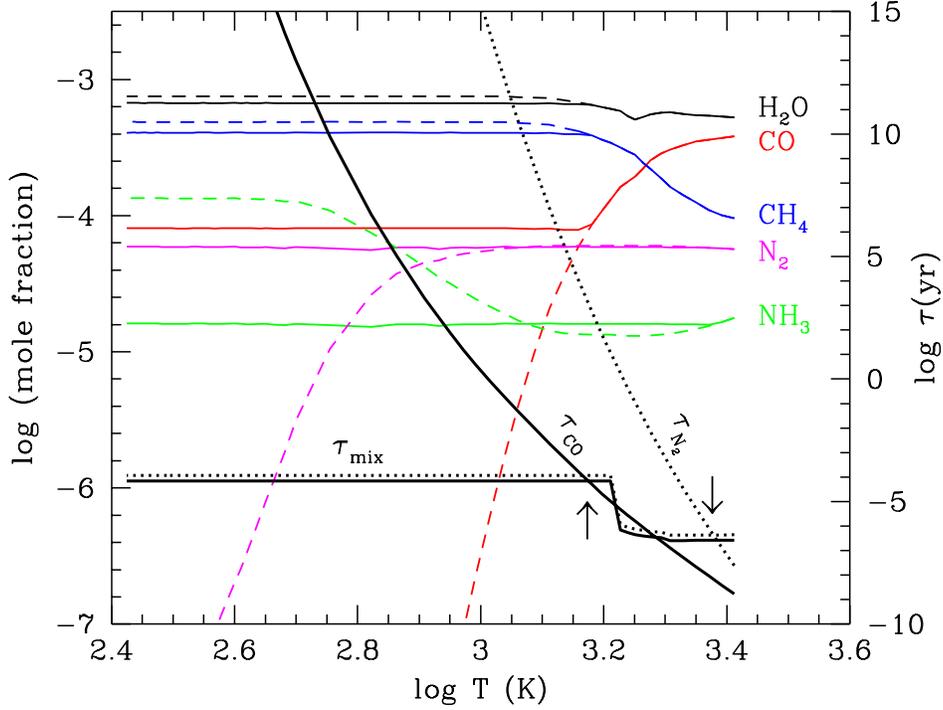}
\caption{Chemical profile of the optimal model atmosphere for Gl~570D: 
$\teff=820\,$K, $\log g=5.23$, [M/H]=0 and $\log 
K_{zz}\,$(cm$^2$/s)=6.2. Mole fractions are shown in equilibrium 
($K_{zz}=0$, dashed curves) and out of equilibrium ($\log 
K_{zz}\,$(cm$^2$/s)=6.2, solid curves) as a function of the temperature 
in the atmosphere (the top of the atmosphere is at left).  Heavy black 
lines show the mixing time scale ($\tau_{\sss {\rm mix}}$) and the time 
scale for the destruction of CO ($\tau_{\sss {\rm CO}}$, solid) and 
N$_2$ ($\tau_{\sss \rm N_2}$, dotted). The mixing time scale is nearly 
discontinuous where the atmosphere becomes convective ($\log T \gtrsim 
3.2$).  The 10--11$\,\mu$m NH$_3$ band is formed at$\log T=2.75--2.90$ 
and the 4.7$\,\mu$m band of CO is formed at $\log T=2.8--2.9$.  [{\it 
See the electronic edition of the Journal for a color version of this 
figure.}]}
\label{fig:chem_Gl570D}
\end{figure}
\clearpage

\begin{figure}
\plotone{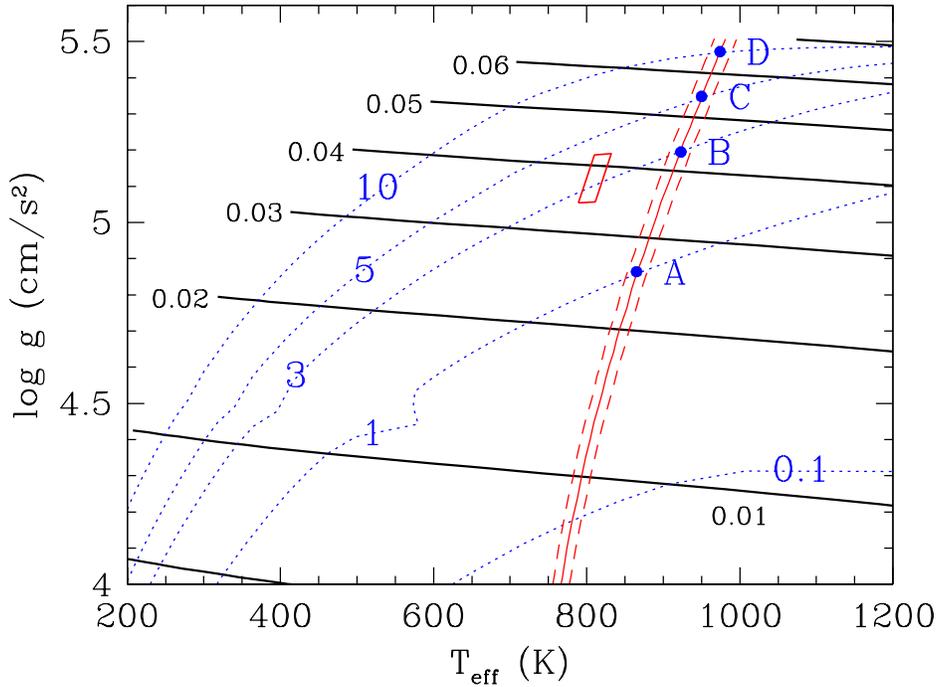}
\caption{$\teff$ and gravity for 2MASS~0937.  Heavy black lines labeled 
with the mass in $M_\odot$ are cloudless brown dwarf evolution tracks 
for [M/H]$=-0.3$. Isochrones (blue dotted lines) are labeled in Gyr.  
The nearly vertical lines (solid and dashed, red) are the locus of 
$(\teff,g)$ points with $\log L/L_\odot = -5.308 \pm 0.027$. The solid 
dots show the four models of Table 2. The range of parameters for 
Gl~570D ([M/H]=0) is shown by the small box centered at 800$\,$K 
\citep{saumon06}. [{\it See the electronic edition of the Journal for a 
color version of this figure.}]}
\label{fig:Teff_g_2M0937}
\end{figure}
\clearpage

\begin{figure}
\plotone{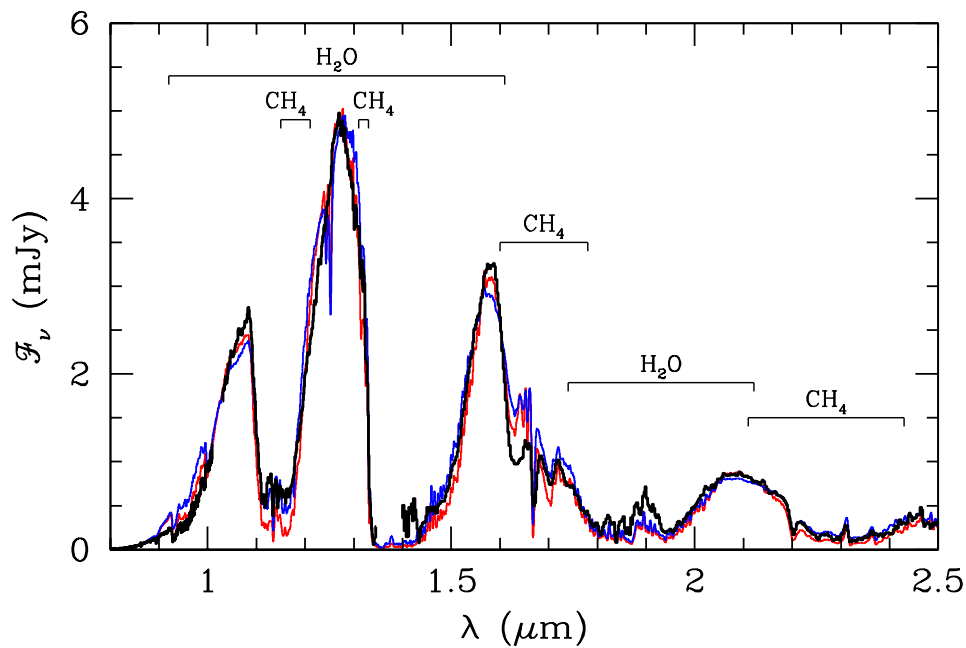}
\caption{Comparison of near infrared synthetic spectra corresponding to 
models A (red) and D (blue) in Table~2 to observed spectrum of 
2MASS~0937 (thick black).  Data are from \citet{bbk06}. The models have 
a [M/H]$=-0.3$, do not include vertical mixing, and are plotted at 
$R=500$.  The main molecular absorbers are indicated. [{\it See the 
electronic edition of the Journal for a color version of this figure.}]}
\label{fig:NIR_2M0937}
\end{figure}
\clearpage

\begin{figure}
\plotone{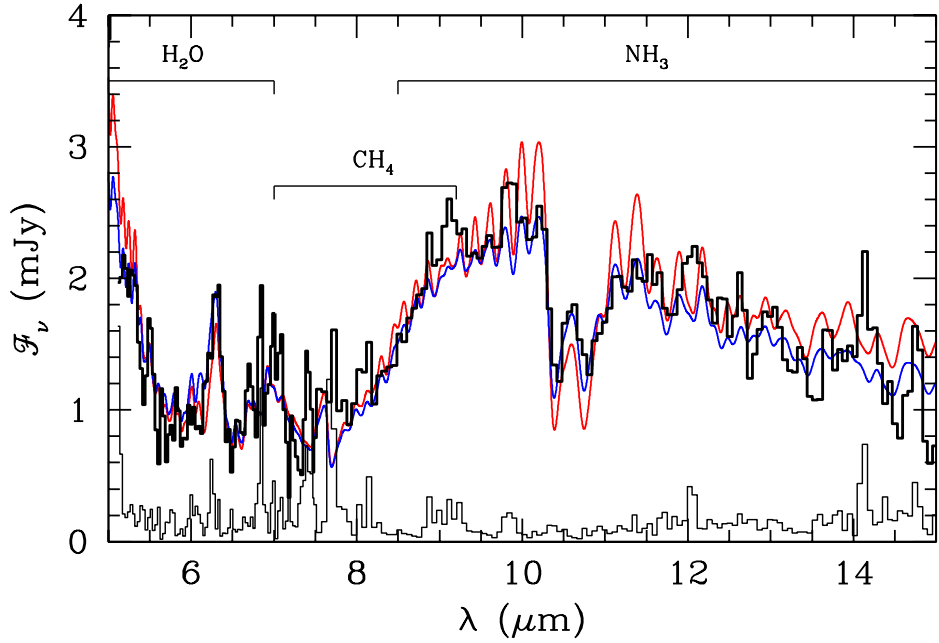}
\caption{Comparison of {\it Spitzer} IRS spectrum of 2MASS~0937 
\citet{cushing06}, black histogram), best fitting model with equilibrium 
chemistry (model~A, Table~2; red), and best fitting non-equilibrium 
model (model~C, Table~2 with $K_{zz}=10^4\,$cm$^2$/s; in blue). The 
lower histogram is the noise spectrum. Flux calibration of the data is 
uncertain to $\pm 3.7$\%.  Model spectra are not renormalized to the 
data.  The models have [M/H]$=-0.3$ and are plotted at $R=120$.  The 
main molecular absorbers are indicated. [{\it See the electronic edition 
of the Journal for a color version of this figure.}]}
\label{fig:MIR_2M0937}
\end{figure}
\clearpage

\begin{figure}
\plotone{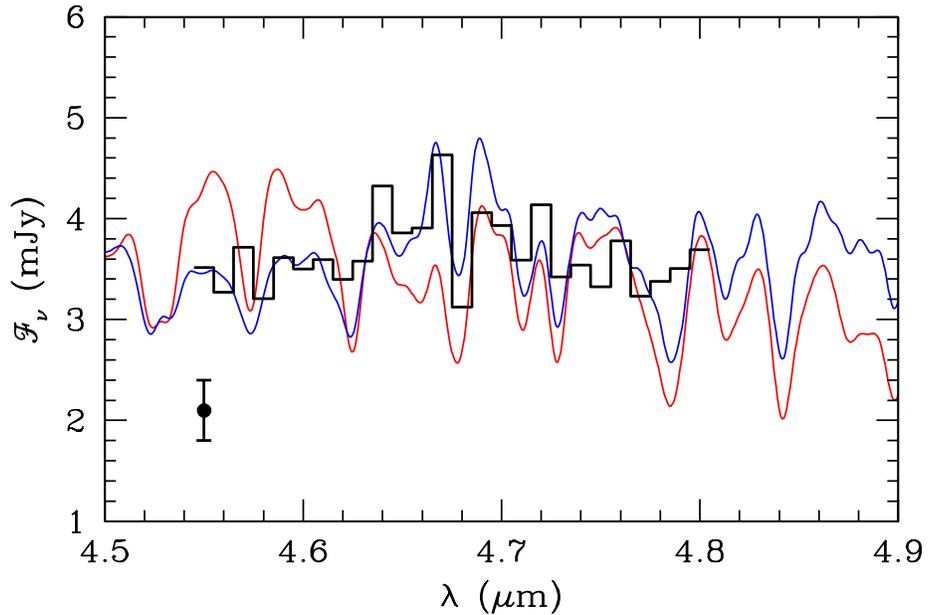}
\caption{Fits to the $M$ band spectrum of 2MASS~0937. Data are the black 
histogram; typical pixel noise level ($\pm 1\sigma$) is shown at lower 
left.  Models with $\teff=950\,$K, $\log g=5.35$, and [M/H]$=-0.3$, 
scaled to minimize residuals, are plotted at $R=500$.  The red curve is 
the equilibrium model ($K_{zz}=0$); the blue curve is the 
non-equilibrium model of Fig.~10 that best fits the 
shape of the spectrum ($\log K_{zz}\,$(cm$^2$/s)=4.3). [{\it See the 
electronic edition of the Journal for a color version of this figure.}]}
\label{fig:Mband_fit_2M0937}
\end{figure}
\clearpage

\begin{figure}
\plotone{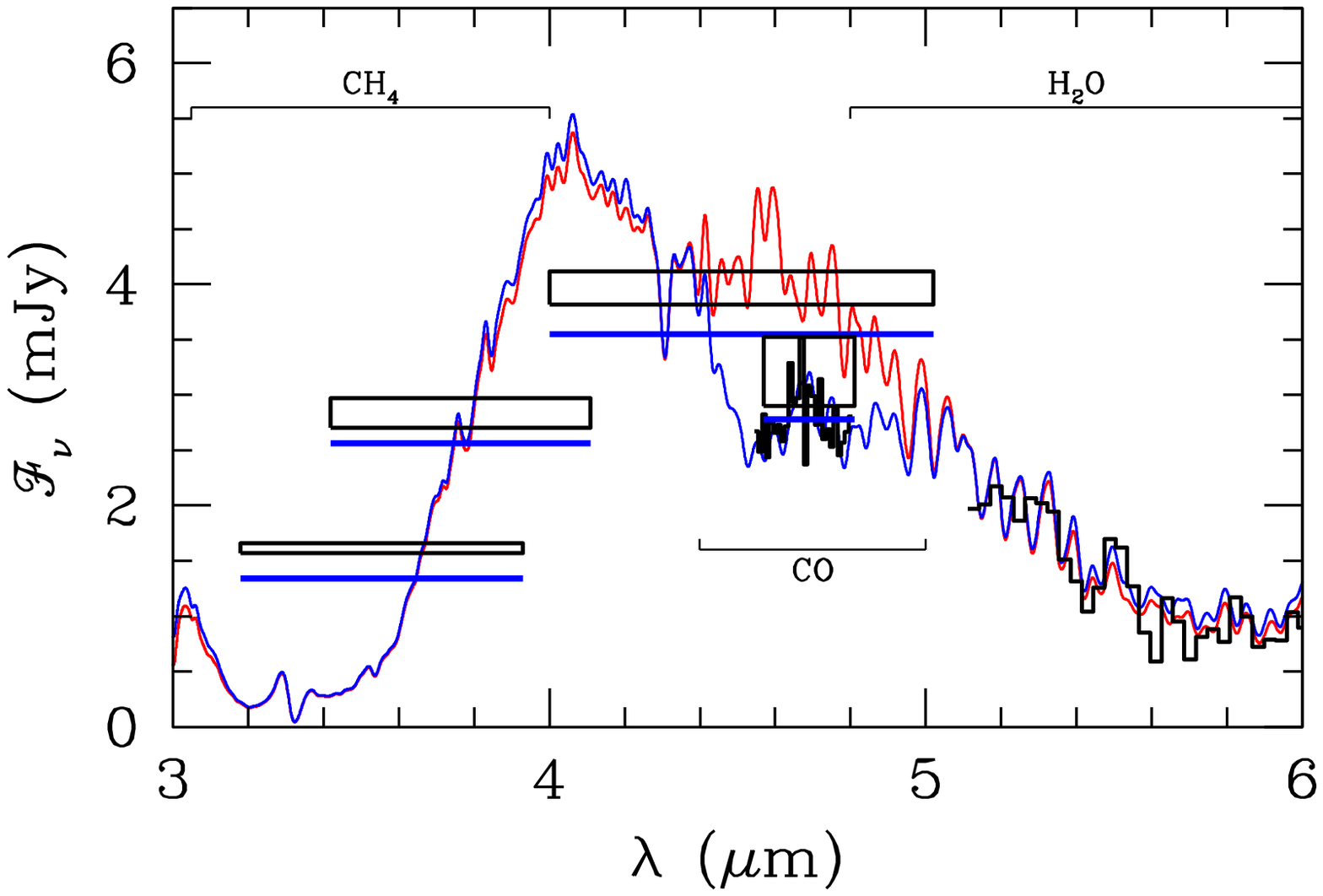}
\caption{3--6~$\mu$m spectral region of 2MASS~0937.  Data are in black.  
Photometric measurements in the IRAC [3.6], MKO $L^\prime$, IRAC [4.5], 
and MKO $M^\prime$ \citep{gol04,patten06} bands are shown by boxes whose 
widths approximate the bandpasses and heights indicate the 
$\pm$1$\sigma$ uncertainty. Two models are shown at $R=200$, both with 
$\teff=950$K, $\log g=5.35$ and [M/H]=$-0.3$.  The red curve shows the 
equilibrium model ($K_{zz}=0$) and the blue curve shows the model that 
best fits the $M$ band spectrum ($\log K_{zz}\,$(cm$^2$/s)=4.3).  The 
synthetic models fluxes over the photometric bandpasses are shown by 
thick blue lines.  The distance is from \citet{perryman97}.  The 
observed spectrum is scaled by a factor of 0.76 to match the 
non-equilibrium model spectrum.  See text. [{\it See the electronic 
edition of the Journal for a color version of this figure.}]}
\label{fig:sp_2M0937_2}
\end{figure}
\clearpage

\begin{figure}
\plotone{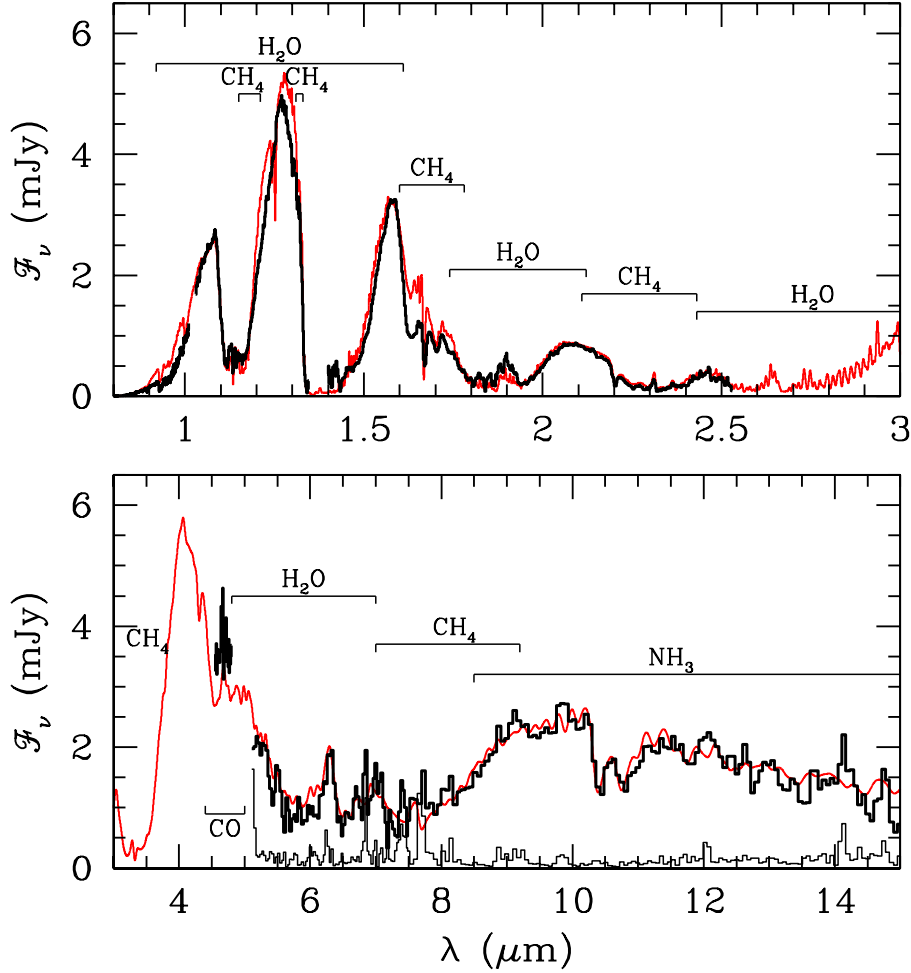}
\caption{Comparison of 2MASS~0937 spectrum (black histograms) with 
non-equilibrium model C (see Table~2) with $\teff=950\,$K, $\log 
g=5.35$, [M/H]=$-0.3$, with $\log K_{zz}\,$(cm$^2$/s)=4.3 (red curves). 
{\it Spitzer} IRS noise spectrum is the lower histogram in the bottom 
panel.  Model spectra are plotted at a resolving power of $R=500$ (upper 
panel) and $R=120$ (lower panel) and are not rescaled to the data.  Main 
molecular absorbers are indicated. [{\it See the electronic edition of 
the Journal for a color version of this figure.}]}
\label{fig:all_2M0937}
\end{figure}
\clearpage

\begin{figure} 
\plotone{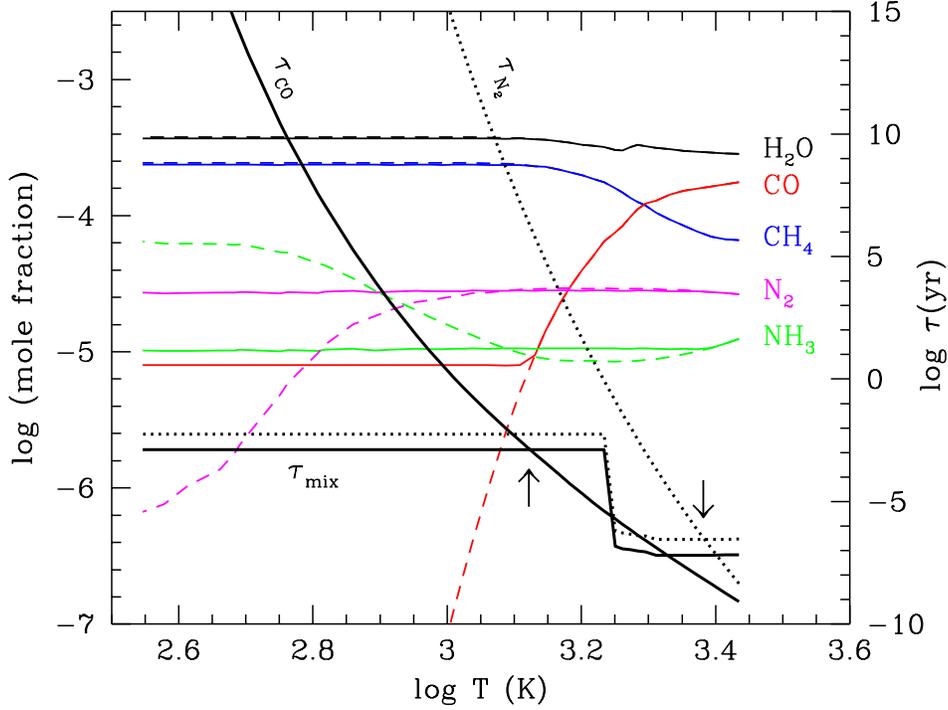} 
\caption{Chemical profile of an optimal model atmosphere for 2MASS~0937 
(model C of Table 2): $\teff=950\,$K, $\log g=5.35$, [M/H]$=-0.3$ and 
$\log K_{zz}\,$(cm$^2$/s)=4.3.  Mole fractions of H$_2$O, CH$_4$, N$_2$, 
NH$_3$ and CO are shown in equilibrium ($K_{zz}=0$, dashed curves) and 
out of equilibrium ($\log K_{zz}\,$(cm$^2$/s)=4.3, solid curves) as a 
function of temperature. Heavy black lines show the mixing time scale 
($\tau_{\sss \rm mix}$) and the time scale for the destruction of CO 
($\tau_{\sss \rm CO}$, solid) and N$_2$ ($\tau_{\sss \rm N_2}$, dotted).  
The mixing time scale is nearly discontinuous where the atmosphere 
becomes convective ($\log T \gtrsim 3.25$).  The 10--11$\,\mu$m NH$_3$ 
band is formed at $\log T=2.80--2.95$ and the 4.7$\,\mu$m band of CO is 
formed at $\log T=2.90--2.98$. [{\it See the electronic edition of the 
Journal for a color version of this figure.}]} 
\label{fig:chem_2M0937} 
\end{figure} 

\clearpage

\begin{deluxetable}{cccc}
\tablewidth{0pt}
\tablecaption{Observing Log}
\tablehead{
\colhead{UT Date} &
\colhead{Source Name} &
\colhead{Exposure (min)} &
\colhead{Calib. star}}
\startdata
20040429 & Gl 570D         &  64 & HIP 78585 \\
20040430 & Gl 570D         &  64 & HIP 78585 \\
20041230 & 2MASS 0937      &  90 & HIP 52959 \\
20050116 & 2MASS 0937      &  84 & HIP 52959 \\
20050126 & 2MASS 0937      & 108 & HIP 52959 \\
20050215 & 2MASS 0937      &  84 & HIP 40843 \\
20050222 & Gl 570D         & 105 & HIP 67945 \\
20050301 & Gl 570D         &  96 & HIP 77610 \\
20050302 & Gl 570D         &  63 & HIP 67945 \\
\enddata
\end{deluxetable}

\tablecolumns{7}
\begin{deluxetable}{cccccccc}
\tablewidth{0pt}
\tablecaption{Range of Physical Parameters for 2MASS
J09373487+2931409\tablenotemark{a}\tablenotemark{b} }
\tablehead{
\colhead{Model} & \colhead{$\teff$}  &  \colhead{$\log g$}  &
\colhead{$\log L/L_\odot$}
 & \colhead{Mass} & \colhead{Radius} & \colhead{Age}  & \colhead{$\log
K_{zz}$\tablenotemark{c}} \\
\colhead{}  & \colhead{(K)}  & \colhead{(cm/s$^2$)} &   \colhead{}  &
\colhead{$(M_J)$}  & \colhead{$(R_\odot)$} & \colhead{(Gyr)} &
\colhead{(cm$^2$/s)}}
\startdata
 A & 865 & 4.86 & $-5.296$ & 27 & 0.0984 &\phs 1.0 \phs   & $-$ \\
 B & 923 & 5.20 & $-5.304$ & 45 & 0.0870 &\phs 3.0 \phn   & $3.2 \pm 0.10$
\\
 C & 950 & 5.35 & $-5.308$ & 57 & 0.0818 &\phs 5.0 \phn   & $3.0 \pm 0.11$
\\
 D & 974 & 5.47 & $-5.312$ & 69 & 0.0778 &\phs 10.0 \phn  & $2.7 \pm 0.14$
\\
\enddata
\tablenotetext{a}{[M/H]=-0.3}
\tablenotetext{b}{The $1\sigma$ uncertainty in the luminosity is $\Delta
\log L/L_\odot=0.027$, corresponding to $\Delta \teff=15\,$K
 and $\Delta R/R_\odot=0.001$ at constant gravity.}
\tablenotetext{c}{For the nominal flux calibration of the data.}
\end{deluxetable}

\end{document}